\documentclass{article}
\usepackage{a4wide}
\newcommand{\scs}{\scriptstyle}
\newcommand{\dys}{\displaystyle}
\renewcommand{\theequation}{\arabic{section}.\arabic{equation}}
\begin{document}
\bibliographystyle{unsrt}
 
\setcounter{page}{0}
\thispagestyle{empty}
\clearpage
  \begin{center}
    {\LARGE\bf
The Quantum Ising Chain\\ with a Generalized Defect  \\ }

   \vspace{2cm}

   {\Large
    Uwe Grimm      \\}

    \vspace{2mm}
  
Physikalisches Institut der Universit\"at Bonn \\
Nu{\ss}allee 12, 5300 Bonn 1, Germany
\vspace{2cm}

Preprint BONN-HE-89-15 (November 1989)

\vspace{0.5cm}

                                                    
\vspace{2cm}

   {\Large
    Abstract\\}

\vspace{5mm}

\end{center}
\begin{quote}
The finite-size scaling properties of the quantum Ising chain with
different types of generalized defects are studied. These not only
mean an alteration of the coupling constant as previously examined,
but an additional arbitrary transformation in the algebra of
observables at one site of the chain.  One can distinguish between two
classes of generalized defects: those which do not affect the
finite-size integrability of the Ising chain, and on the other hand
those that destroy this property.  In this context, finite-size
integrability is always understood as a synonym for the possibility to
write the Hamiltonian of the finite chain as a bilinear expression in
fermionic operators by means of a Jordan-Wigner transformation.
Concerning the first type of defect, an exact solution for the scaling
spectrum is obtained for the most universal defect that preserves the
global ${\bf Z}_2$ symmetry of the chain.  It is shown that in the
continuum limit this yields the same result as for one properly chosen
`ordinary' defect, that is changing the coupling constant only, and
thus the finite-size scaling spectra can be described by irreps of a
shifted $u(1)$ Kac-Moody algebra.  The other type of defect is
examined by means of numerical finite-size calculations.  In contrast
to the first case, these suggest a non-continuous dependence of the
scaling dimensions on the defect parameters.  A conjecture for the
operator content involving only one primary field of a Virasoro
algebra with central charge $c\! =\! \frac{1}{2}$ is given.
\end{quote}
\vfill

(This is a reprint version of a paper that appeared in
Nucl.\ Phys.\ B {\bf 340} (1990) 633--658.)
\clearpage

\section{Introduction}
\setcounter{equation}{0}
\pagestyle{plain}
 
The one-dimensional quantum Ising chain, the Hamiltonian limit of the
classical two-dimensional Ising model \cite{ising} obtained via the
transfer-matrix approach (see \cite{kog,bax} and references contained
therein) belongs to the most frequently investigated and best
understood systems in statistical physics.  The main cause of this is
the fact that the Ising model (without an external magnetic field,
however) admits an analytic solution \cite{ons} and on top of this
undergoes a continuous phase transition at a finite temperature. It is
known \cite{lsm} that at criticality the Ising model describes the
field theory of a free massless Majorana fermion, which is a conformal
field theory with the central charge $c$ of the Virasoro algebra (VA)
being \mbox{$c\! = \! \frac{1}{2}$}.
 
After the importance of the conformal structure \cite{bpz,fqs,ca} of
spin chains at the critical point had been recognized, the question
arose how defects such that the system stays critical will affect
these properties.  Clearly one started the investigation of this
problem with the simplest example, that is with the Ising quantum
chain. The defects in this case were introduced by changing the
coupling constant of one nearest-neighbour coupling corresponding to a
change of the coupling constant on a half-infinite line in the
two-dimensional classical model. The system in this case stays
critical \cite{tu} and part of the conformal structure is maintained
\cite{hp,hps,bcs,schl}.
 
It is the aim of this article to give a generalization of the case of
one localized defect. This is done by performing an arbitrary
transformation in the algebra of observables at one site of the
chain. In other words, this means that one replaces the Pauli matrix
$\sigma^{x}$ in one coupling term of the $N$-site Hamiltonian $H$ of
the (ferromagnetic) critical quantum Ising chain with periodic
boundary conditions
\begin{equation} 
H = - \frac{1}{2} \sum_{j=1}^{N} \left[
\sigma_{j}^{z} + \sigma_{j}^{x} \sigma_{j+1}^{x} \right] \; \; \; , \;
\; \; \sigma^{x}_{N+1} = \sigma^{x}_{1} \; \; ,
\label{I:H}
\end{equation} 
by a general hermitian \mbox{$2\! \times \! 2$} matrix, therefore one
has four real parameters $e_{1},e_{x},e_{y},$ and $e_{z}$ that
determine the generalized defect by
\begin{equation} 
\sigma^{x}_{N+1}
= e_{1}{\bf 1} + e_{x}\sigma_{1}^{x} + e_{y}\sigma_{1}^{y} +
e_{z}\sigma_{1}^{z} \; \; ,
\label{I:gd} 
\end{equation}
where ${\bf 1}$ denotes the \mbox{$2\! \times \! 2$} unit matrix and
an `ordinary' defect corresponds to \mbox{$e_{1}\! =\! e_{y}\! =\!
e_{z}\! = \! 0$}.
 
The first results for the Ising quantum chain with `ordinary' defects
were obtained by detailed numerical calculation \cite{hp}, but
recently \cite{hps,schl} the exact solution for an arbitrary number of
isolated defects has been found. The algebraic structure of the
spectrum for a fixed number of defects at commensurate distances has
been examined in detail in \cite{bcs} (see also \cite{schl}).  The
results of these investigations for the case of one defect, where the
spectrum generating algebra turns out to be a supersymmetric extension
(the \mbox{$\frac{p}{q} \! = \! \frac{1}{2}$} higher symmetry algebra
\cite{bcr}) of a so-called `shifted' $u(1)$ Kac-Moody algebra (will be
abbreviated by KMA in the sequel), are summarized in the subsequent
section.  Even more recently, another type of defect (a so-called
`extended' defect), where the coupling constant is altered on a whole
fraction of the chain, has been studied \cite{haye} and a similar
algebraic structure as above has been obtained.
 
In contrast to the case of an `ordinary' defect, a generalized defect
as defined above might or might not destroy the finite-size
integrability property of the chain and one expects a quite different
behaviour for various types of generalized defects. Here and in what
follows, finite-size integrability always means that one is able to
write the Hamiltonian of the finite chain as a bilinear expression in
fermionic operators via a Jordan-Wigner transformation \cite{lsm}.  A
generic example consisting of a two-parameter family of defects for
each of this two classes is examined by analytical respective
numerical methods. In the case where the chain stays finite-size
integrable the results of the exact solution resemble those of one
`ordinary' defect, which is contained therein for a special choice of
the parameters.  In particular, the scaling dimensions depend
continuously on the parameters of the defects, whereas in the other
case the numerical results favour a non-continuous behaviour. The
operator content in the latter case is conjectured to be given by a
single irrep of a VA with central charge \mbox{$c\! =\! \frac{1}{2}$}
for fixed parameters of the defect.  This article restricts to the
quantum chain point of view, physical implications for the
two-dimensional classical system --- if any -- are not discussed in
this place.
 
The paper is organized as follows.  In the subsequent section the
previously examined boundary conditions and `ordinary' defects
together with the `generalized' defects (\ref{I:gd}) are introduced
more accurately.  A short summary of the formerly obtained results is
given.  This includes a presentation of the character functions of
unitary highest weight irreps of the VA with central charge \mbox{$c\!
=\! \frac{1}{2}$} and of a shifted $u(1)$ KMA.  The generalized
defects are then divided into two groups: one type that leaves the
quantum chain finite-size integrable and the other that destroys this
property.
 
The third section contains an exact solution for the spectrum of the
chain with an finite-size integrable type of defect that maintains the
global ${\bf Z}_2$ symmetry of the Ising chain corresponding to
\mbox{$e_{1}\! = \! e_{z}\! = \! 0$} in Eq.~(\ref{I:gd}).  This is
done employing the methods developed by Lieb, Schultz and Mattis
\cite{lsm} (see also \cite{hps,schl}) in a slightly generalized
form. The operator content is given in terms of character functions of
irreducible representations (irreps) of a shifted $u(1)$ KMA and in
fact turns out to have the same form as for one `ordinary' defect
\cite{bcs}, whereas the behaviour on finite lattices appears to be
different in general.
 
In the forth part an example for a two-parameter class of finite-size
non-integrable defects that corresponds to taking \mbox{$e_{1}\! =\!
e_{y}\! = \! 0$} in Eq.~(\ref{I:gd}) is studied.  This can be done
through the application of numerical methods only. The effect of this
type of defect on the finite-size scaling limit appears to differ
qualitatively from the results of Sec.~3. The numerical data indicate
that the dependence of the critical exponents on the parameters of the
defect is not a continuous one.  A conjecture for the operator content
based on finite-size calculations for chains with a length of up to
sixteen sites is presented. It can be expressed in terms of irreps of
the VA with central charge $c\! =\! \frac{1}{2}$ involving only one
primary field in each case.
 
The main results of this paper are summarized in the concluding
Sec.~5.  On top of this some questions that remained open are
discussed in this place.  The two Appendices consist of proofs
concerning parts of the exact solution in Sec.~3 that had been omitted
therein for the sake of the reader.

\section{`Ordinary' and `generalized' defects}
\setcounter{equation}{0}
 
In terms of the Pauli matrices $\sigma^{x}$, $\sigma^{y}$, and
$\sigma^{z}$, the $N$-site ferromagnetic quantum Ising Hamiltonian
reads
\begin{equation} 
H = - \frac{1}{2} \sum_{j=1}^{N} \left[ \sigma_{j}^{z} +
\lambda \sigma_{j}^{x} \sigma_{j+1}^{x} \right] \; \; ,
\label{e:H}
\end{equation} 
where $\lambda$ plays the role of the inverse temperature and the
meaning of $\sigma_{N+1}^{x}$ has to be fixed by an appropriate
boundary condition (BC). In this article only those BCs (or `defects')
will be considered that can be expressed in this way and where in
addition $\sigma_{N+1}^{x}$ depends only on variables that live on the
first site. Thus in Eq.~(\ref{e:H}) only (at most) two adjacent sites
are coupled directly.  The Hamiltonian (\ref{e:H}) with a proper
choice of the boundary term enjoys a global ${\bf
Z}_{2}$-invariance. It commutes with the `charge' operator $\hat{Q}$
defined by
\begin{equation} 
\begin{array}{cccc}
\hat{Q} = \dys \prod_{j=1}^{N} 
\sigma_{j}^{z} & , & \hat{Q}^{2} = {\bf 1} & ,
\label{e:Q}
\end{array} 
\end{equation} 
where ${\bf 1}$ denotes the identity operator. One can therefore
divide the spectrum of $H$ (\ref{e:H}) into two sectors corresponding
to an eigenvalue $(-1)^q$ of $\hat{Q}$ with \mbox{$q \!  \in \! \{ 0,1
\} $}.  The system described by $H$ (\ref{e:H}) is the Hamiltonian
limit of the classical two-dimensional Ising model that exhibits a
continuous phase transition at $\lambda \! = \! 1$ and the finite-size
scaling limit of the spectrum is known for various different BCs, such
as toroidal, free, fixed and mixed BCs \cite{bg,ca,gr} and for the
case of an arbitrary number of isolated (`ordinary') defects
\cite{hps,bcs}.  Before starting with the subject of this paper and
defining new possible choices of BCs, let us have a look to the
formerly used BCs and the results that have been obtained for these.
 
One already mentioned group of BCs is the so-called `one defect'
\cite{hp} (in the sequel the notion `ordinary' defect will be used to
avoid confusion with the generalized defects) of strength $e_{x}$,
where one defines $\sigma_{N+1}^{x}$ to be (cf. Eq.(\ref{I:gd}))
\begin{equation}
\sigma_{N+1}^{x} = e_{x} \cdot \sigma_{1}^{x}
\label{e:k}
\end{equation}
with \mbox{$e_{x} \! \in \! {\bf R}$}, which in the two-dimensional
model corresponds to a change of the coupling constant on a
half-infinite line \cite{hp}.  For special values of the parameter
$e_{x}$, this includes most of the previously examined possible
boundary terms.  There are two distinguished choices for
$\sigma_{N+1}^{x}$ in (\ref{e:H}) that correspond to periodic and
antiperiodic BCs given by \mbox{$e_{x} \! = \! 1$} and \mbox{$e_{x} \!
= \! -1$}, respectively.  These are the sole BCs that allow the
definition of a translation operator that commutes with $H$
(\ref{e:H}) and therefore are called toroidal BCs (see \cite{ch}).
Free BCs occur in this context as the case \mbox{$e_{x}\! = \!
0$}. The also previously examined so-called fixed \cite{bg} BCs ---
which by performing a duality transformation can be shown to yield the
same results as free BCs --- as well as mixed BCs \cite{bg} cannot be
written in this simple form, they are obtained by attaching constant
terms to the two ends, respectively to one end (leaving the other end
free), of the chain.
 
All the above-mentioned BCs (or defects) have in common that the
Hamiltonian $H$ (\ref{e:H}) commutes with the charge operator
$\hat{Q}$ (\ref{e:Q}) and beyond it can be written as a bilinear
expression in fermionic operators.  The last property ensures that the
system stays finite-size integrable in each case.  Therefore, the
finite-size scaling limit spectra are known exactly for all these
cases. In general, they can be expressed in terms of irreps of a
`shifted' $u(1)$ KMA \cite{bcr,mb} that contains a VA with central
extension \mbox{$c \! = \! 1$}.  For the special cases of free (as
also for fixed and mixed) BCs it is also possible to use one VA with
central charge $c\! = \! \frac{1}{2}$ to describe the spectrum in the
finite-size scaling limit (\cite{ca,gr}, see also Sec.~4), whereas for
toroidal boundary conditions one has two commuting VAs with \mbox{$c
\! = \! \frac{1}{2}$} \cite{ca} due to translational invariance.
 
To be more precise, let me reformulate the contents of the last
paragraph in a more accurate way.  As already mentioned, the operator
content for free and toroidal BCs can be expressed \cite{ca,gr} in
terms of unitary irreps of one respective two commuting VAs with $c\!
= \! \frac{1}{2}$.  For general central extension $c$, the Virasoro
generators $L_{m}\,$,\mbox{$\; m\! \in \! {\bf Z}$}, fulfill the
commutation relations
\begin{equation}
\left[ L_{m}, L_{n} \right] = 
(m-n) L_{m+n} + \frac{c}{12} m (m^2-1) \delta_{m+n,0}
\label{e:VA}
\end{equation}
with $m,n \in {\bf Z}$.  The VA with central charge $c\! = \!
\frac{1}{2}$ has only three unitary irreps \cite{fqs} corresponding to
a highest weight \mbox{$\Delta \! \in \! \{ 0,\frac{1}{16},\frac{1}{2}
\} $}. The character functions \mbox{$\chi_{\Delta}(z) = tr
(z^{L_{0}})$} in this case are given by \cite{rc,mbt}
\begin{equation} 
\begin{array}{lclcl}
\chi_{0}(z)            & = & \dys{\sum_{n \in {\bf Z}}} 
                             \left( z^{(12n^{2}+n)} - z^{(12n^{2}+7n+1)} 
                             \right) \cdot \Pi_{V}(z)
                       & = & \dys{\sum_{n \in {\bf Z}}} z^{4n^2+n} 
                             \cdot \Pi_{V}(z^2) \vspace{4mm} \\
\chi_{\frac{1}{16}}(z) & = & z^{\frac{1}{16}} \cdot 
                             \dys{\sum_{n \in {\bf Z}}}
\left( z^{(12n^{2}+2n)} - z^{(12n^{2}+14n+4)} \right) \cdot \Pi_{V}(z)
                       & = & z^{\frac{1}{16}} \cdot 
                             \dys{\prod_{m=1}^{\infty}}
\left( 1-z^{2m} \right) \cdot \Pi_{V}(z) \vspace{4mm} \\
\chi_{\frac{1}{2}}(z)  & = & z^{\frac{1}{2}} \cdot 
                             \dys{\sum_{n \in {\bf Z}}} 
                             \left( z^{(12n^{2}+5n)} - 
                             z^{(12n^{2}+11n+2)} \right)
\cdot \Pi_{V}(z)       & = & z^{\frac{1}{2}} \cdot 
                             \dys{\sum_{n \in {\bf Z}}} z^{4n^2+3n} 
                             \cdot \Pi_{V}(z^2) \; \; .
\end{array} \label{e:cf}
\end{equation}
Here, $\Pi_{V}(z)$ denotes the function generating the number of
partitions, which is given as an infinite product by the expression
\begin{equation}
\Pi_{V}(z) = \prod_{m=1}^{\infty} \left( \frac{1}{1-z^{m}} \right) \; \; .
\label{e:pv}
\end{equation}
The partition functions ${\cal T}_{q}^{\pm}(z,\bar{z})$ resp. ${\cal
F}_{q}(z)$ for the finite-size scaling spectra of the Hamiltonian
$H$~(\ref{e:H}) in the charge sector $q$, \mbox{$\; q \! \in \!
\{0,1\}$} are defined by \cite{ca}
\begin{equation} 
\begin{array}{lcl}
{\cal T}_{q}^{\pm}(z,\bar{z}) & = & 
\dys{\lim_{N \rightarrow \infty}} \dys{\sum_{k,P}}
z^{\frac{1}{2} \left( \overline{T}_{q;k}^{\,\pm}(P,N) + P \right) }
\bar{z}^{\frac{1}{2} \left( \overline{T}_{q;k}^{\,\pm}(P,N) - P \right) }
\vspace{4mm} \\
\overline{T}_{q;k}^{\,\pm}(P,N) & = & 
\frac{N}{2\pi} \left( E_{q;k}^{\pm}(P,N) - E_{0;0}^{+}(0,N) \right)
\end{array} \label{e:T} 
\end{equation}
for toroidal BCs and \cite{ca,gr}
\begin{equation} 
\begin{array}{lcl}
{\cal F}_{q}(z)       & = & \dys{\lim_{N \rightarrow \infty}} 
\dys{\sum_{k}} z^{\overline{F}_{q;k}(N)} \vspace{4mm} \\
\overline{F}_{q;k}(N) & = & \frac{N}{\pi} \left( E_{q;k}^{f}(N) - 
E_{0;0}^{f}(N) \right)
\end{array} \label{e:F} 
\end{equation}
for free BCs. Here, $E_{q;k}^{\pm}(P,N)$ denotes the $k^{th}$
eigenvalue of the Hamiltonian (\ref{e:H}) with periodic ($+$) ---
resp. antiperiodic ($-$) --- BCs in the charge sector $q$ and with the
momentum (eigenvalue of the translation operator) $P$ and the sum
covers the whole spectrum of $H$ (\ref{e:H}).  The analogue is true
for the energies $E_{q;k}^{f}(N)$ for free BCs, except that in this
case there is no translation operator commuting with $H$~(\ref{e:H}).
The partition functions (\ref{e:T}) and (\ref{e:F}) are then given in
terms of the character functions tabulated in Eq.~(\ref{e:cf}) by
\cite{ca}
\begin{equation} 
\begin{array}{lcl}
{\cal T}_{0}^{+}(z,\bar{z}) & = & \chi_{0}(z) \chi_{0}(\bar{z})
+ \chi_{\frac{1}{2}}(z) \chi_{\frac{1}{2}}(\bar{z}) \vspace{4mm} \\
{\cal T}_{1}^{-}(z,\bar{z}) & = & \chi_{\frac{1}{2}}(z) \chi_{0}(\bar{z})
+ \chi_{0}(z) \chi_{\frac{1}{2}}(\bar{z}) \vspace{4mm} \\
{\cal T}_{0}^{-}(z,\bar{z}) & = & {\cal T}_{1}^{+}(z,\bar{z}) 
\; \; = \; \; \chi_{\frac{1}{16}}(z) \chi_{\frac{1}{16}}(\bar{z})
\end{array} \label{e:oct} 
\end{equation}
and by \cite{ca,gr}
\begin{equation} 
\begin{array}{ccccccc} {\cal F}_{0}(z) & = & \chi_{0}(z) & , & 
{\cal F}_{1}(z) & = & \chi_{\frac{1}{2}}(z) \; \; .  
\end{array} \label{e:ocf} 
\end{equation}
For completeness, let me state the results for fixed and mixed BCs.
The Hamiltonians for these BCs can be summarized in the form
\begin{equation}
H(f_{1},f_{N}) = -\frac{1}{2} \left( \sum_{j=1}^{N-1} 
\left[ \sigma^{x}_{j} \sigma^{x}_{j+1} + \sigma^{z}_{j} \right]
                 + \sigma^{z}_{N} + f_{1} \cdot \sigma^{x}_{1} + 
f_{N} \cdot \sigma^{x}_{N} \right) 
\label{e:hfm}
\end{equation}
with \mbox{$f_{1},f_{N} \in \{ 1,0,-1 \} \; $}.  Here, the different
sectors of fixed BCs correspond to the choices \mbox{$f_{1}\! =\! \pm
1$} and \mbox{$f_{N}\! =\! \pm 1$}, whereas mixed BCs are defined by
\mbox{$f_{1}\! =\! 0$} and \mbox{$f_{N}\! =\! \pm 1$} (or,
equivalently \mbox{$f_{N}\! =\! 0$} and \mbox{$f_{1}\! =\! \pm 1$}).
The partition functions for the finite-size scaling spectra in each
case are defined relative to the ground-state of the chain with free
BCs as in Eq.~(\ref{e:F}). For fixed BCs \cite{ca} , they are given by
$\chi_{0}(z)$ for equal fixed boundary spins (that is for $H(1,1)$ and
$H(-1,-1)\;$) and by $\chi_{\frac{1}{2}}(z)$ for different fixed
boundary spins (that is for $H(1,-1)$ and $H(1,-1)\;$) at the two ends
of the chain, with these two BCs being related to the two charge
sectors of free BCs by means of a duality transformation.  In the case
of mixed BCs \cite{ca} , one obtains $\chi_{\frac{1}{16}}(z)$ for the
properly defined (that is, a non-universal surface term has to be
taken into account) partition function of the finite-size scaling
spectrum of $H(0,\pm 1)\;$ (\ref{e:hfm}).
 
Now, let me come back to the general case of one `ordinary' defect of
strength $e_{x}$~(\ref{e:k}).  Instead of the
Hamiltonian~$H(e_{x})$~(\ref{e:H}) with the defect $e_{x}$~(\ref{e:k})
one considers the mixed-sector Hamiltonian $\tilde{H}(e_{x})$
\cite{bcs}
\begin{equation}
\tilde{H}(\kappa) = H(e_{x}) \cdot P_{0} + H(-e_{x}) \cdot P_{1} \; \; ,
\label{e:Ht}
\end{equation}
where $P_{q}$ denotes the projection operator
\begin{equation}
P_{q} = \frac{1}{2} \cdot \left( \hat{Q} + (-1)^{q}) \right)
\label{e:P}
\end{equation}
onto the eigenspace belonging to an eigenvalue $(-1)^{q}$ of the
charge operator $\hat{Q}$ (\ref{e:Q}), \mbox{$q \! \in \! \{ 0,1\}$}.
In this way one avoids the appearance of a non-local number operator
\cite{bcs} in the Hamiltonian. Of course, one can reconstruct the
spectrum of $H$ (\ref{e:H}) from the knowledge of the spectrum of
$\tilde{H}$ (\ref{e:Ht}), thus no information has been lost.  The
partition function \mbox{${\cal E}(e_{x},z)$} of the finite-size
scaling spectra (relative to the ground-state energy for periodic BCs
\mbox{($e_{x}\! = \! 1$)}, cf. \cite{bcs}) for the chain with one
defect of strength $e_{x}$ is then defined by (cf. Eqs.~(\ref{e:T})
and (\ref{e:F}))
\begin{equation} 
\begin{array}{ccl}
{\cal E}(e_{x},z) & = & \dys{\lim_{N \rightarrow \infty}}
\dys{\sum_{k}} z^{\overline{E}_{k}(e_{x} ,N)} \vspace{4mm} \\
\overline{E}_{k}(e_{x} ,N) & = & \dys{\frac{N}{2\pi}} \cdot \left(
E_{k}(e_{x},N) + A_{1}(e_{x}) -E_{0}(1,N) \right) \; \; , \end{array}
\label{e:ss}
\end{equation}
where $E_{k}(e_{x},N)$ denotes the $k^{th}$ eigenvalue of the $N$-site
Hamiltonian $\tilde{H}$ (\ref{e:Ht}) with the defect $e_{x}$
(\ref{e:k}). The non-universal surface-energy term $A_{1}(e_{x})$ that
occurs in the expansion of the lowest eigenvalue $E_{0}(e_{x})$ of the
Hamiltonian~(\ref{e:Ht}) with the defect $e_{x}$ \cite{bcn,aff}
\begin{equation}
-E_{0}(e_{x}) = A_{0} \cdot N + A_{1}(e_{x}) + A_{2}(e_{x}) 
\cdot N^{-1} + o(N^{-1})
\label{e:gse}
\end{equation}
has been subtracted (the bulk term \mbox{$A_{0}\! = \! \frac{2}{\pi}$}
of course is not affected by a defect that alters only one term in the
Hamiltonian~(\ref{e:Ht})).  The dependence of this surface term on
$e_{x}$ is known analytically \cite{bcs} and it is given by
\begin{equation}
A_{1}(e_{x}) = \frac{1+e_{x}^{2}}{\pi e_{x}} 
\arctan (e_{x} ) - \frac{1}{2} \; \; ,
\label{e:A1}
\end{equation}
thence in particular \mbox{$A_{1}(0) = \frac{1}{\pi} - \frac{1}{2}$}
and \mbox{$A_{1}(\pm 1)\! = \! 0$}, as necessary for toroidal BCs.
Furthermore, the values of $A_{2}(e_{x})$ in these cases are given by
\cite{ca} \mbox{$A_{2}(1)\! = \! \pi/12\;$}, \mbox{$A_{2}(-1)\! = \!
-\pi/6\;$}, and \mbox{$A_{2}(0)\! = \! \pi/48\;$}.
 
The finite-size scaling limit spectra ${\cal E}(e_{x},z)$ (\ref{e:ss})
can be expressed \cite{bcs} in terms of character functions of unitary
irreps of a shifted $u(1)$ KMA.  For this purpose define the $u(1)$
KMA by the commutation rules
\begin{equation} 
\begin{array}{ccl}
\left[ T_{m} ,T_{n}  \right] &  = & m \cdot \delta_{m+n,0} \vspace{4mm} \\
\left[ T_{m} ,L_{n}  \right] &  = & m \cdot T_{m+n}
\end{array} \label{e:U1}
\end{equation}
with $m,n\! \in \! {\bf Z}\;$, where the operators $L_{m}$ generate a
VA (\ref{e:VA}) with the central charge \mbox{$c \! = \! 1$}. The
$L_{m}$ result from the $T_{m}$ by Sugawaras construction \cite{go}
through \mbox{$L_{m} = \frac{1}{2} \dys{\sum_{r \in {\bf Z}}} :
T_{m-r} T_{r} : \; $}, where the colons denote normal ordering.  One
then introduces a shift $\varphi$ by means of the automorphism
\begin{equation} 
\begin{array}{ccccl}
T_{m} & \longmapsto & \tilde{T}_{m}(\varphi) & = & 
T_{m} + \varphi \cdot \delta_{m,0} \vspace{4mm} \\
L_{m} & \longmapsto & \tilde{L}_{m}(\varphi) & = & 
L_{m} + \varphi \cdot T_{m} + \frac{1}{2} \varphi^{2} \cdot \delta_{m,0}
\end{array} \label{e:shift}
\end{equation}
that leaves the commutation relations~(\ref{e:U1})~and~(\ref{e:VA})
invariant.  The unitary highest weight irreps of this algebra have the
character functions
\begin{equation}
\chi_{t}(z,y) = tr \left( z^{\tilde{L}_{0}(\varphi)} 
\cdot y^{\tilde{T}_{0}(\varphi)} \right)
= z^{\frac{1}{2}(t+\varphi)^{2}} \cdot \Pi_{V}(z) 
\cdot y^{t+\varphi} \; \; ,
\label{e:cfU} 
\end{equation}
indexed by a single real number \mbox{$t\! \in \! {\bf R}$} which is
the eigenvalue of $T_{0}$.  Here, $\Pi_{V}(z)$ is the function
previously defined in Eq.~(\ref{e:pv}).  In the physical application
one is interested in the eigenvalue of $\tilde{L}_{0}(\varphi)$ in
Eq.~(\ref{e:cfU}) only, therefore one uses the character functions
setting \mbox{$y\! = \! 1$} and thereby leaving the actual value of
$t$ (there are in general two different ones) unspecified.
 
The partition functions ${\cal E}(e_{x},z)$ (\ref{e:ss}) of the
finite-size scaling spectra for the Hamiltonian $\tilde{H}$
(\ref{e:Ht}) with one defect of strength $e_{x}$ are then given by
\begin{equation}
{\cal E}(e_{x},z)  =  tr \left( z^{\tilde{L}_{0}(\varphi(e_{x}))} \right)
= \sum_{n \in {\bf Z}} z^{\frac{1}{2} 
\left( n + \varphi(e_{x}) \right)^{2}} \cdot \Pi_{V}(z)
\label{e:ek}
\end{equation}
with $\varphi(e_{x})$ determined by \cite{hps,bcs}
\begin{equation}
\varphi(e_{x}) = \frac{1}{4} - \frac{1}{\pi} \arctan (e_{x}) \; \; .
\label{e:phi}
\end{equation}
One recognizes that this is an infinite sum of characters of irreps of
the $\varphi$ - shifted $u(1)$ KMA (\ref{e:shift}). Equivalently one
could interpret this as a single irrep of the shifted $\frac{p}{q} \!
= \! \frac{1}{2}$ higher symmetry algebra \cite{bcr}.
 
After this short r\'{e}sum\'{e} of known facts about the Ising quantum
chain let me come back to the purpose of this article.  As already
defined in Eq.~(\ref{I:gd}) of the preceding section, the notion of a
`generalized' defect means that one introduces a boundary term by
defining $\sigma_{N+1}^{x}$ in Eq.~(\ref{e:H}) to be any expression
subject to the sole requirement that it involves only variables living
on the first site of the chain (and, of course, that the Hamiltonian
(\ref{e:H}) stays hermitian).  The most universal term compatible with
this condition will obviously be of the form~(\ref{I:gd}), thus one
has four real parameters that determine the defect.  It is clear that
all those defects with $e_{1}$ or $e_{z}$ different from zero will
destroy the global ${\bf Z}_2$ symmetry of the chain. Therefore, one
can distinguish between those defects that preserve the symmetry and
those, which break it (for more complicated models that the Ising
model, there might be different degrees of symmetry breaking by
defects that one would have to consider separately). More important
for the analysis is the question, if the defect will influence the
finite-size integrability property of the chain, thus this will be the
classification of the `generalized' defects that will be used in what
follows.
 
It is not the aim of this paper to exhaust all possible choices in
Eq.~(\ref{I:gd}), but rather to restrict to the examination of two
generic cases, each consisting of two-parameter families of
generalized defects.  In the subsequent section, the most general
defect preserving the global ${\bf Z}_{2}$ symmetry of the chain, that
is \mbox{$e_{1}\! = \! e_{z}\! = \! 0$} in Eq. (\ref{I:gd}), will be
studied. This always leads to an finite-size integrable Hamiltonian
(however, it is not the most general type of defect doing this,
another example for defects that share this property are for instance
given by taking in Eq.~(\ref{I:gd}) \mbox{$e_{x}\! = \! e_{y}\! = \!
e_{z}\! = \! 0$} and $e_{1}$ arbitrary, which for $e_{1}\! \neq \! 0$
break the ${\bf Z}_{2}$ symmetry).  A completely different type of
defect is considered in Sec.~4. This is given by letting in
Eq.~(\ref{I:gd}) $e_{1}$ and $e_{y}$ be equal to zero and the other
two coefficients unspecified.  The defects of this type have the
property to destroy both the global symmetry and the finite-size
integrability of the chain as long as $e_{z}$ does not vanish.

\section{The exact solution for one finite-size integrable 
generalized defect}
\setcounter{equation}{0}

Consider the two-parameter family of Hamiltonians
\begin{equation}
H(\alpha ,\phi ) = -\frac{1}{2} \left( \sum_{j=1}^{N-1} 
                 \left[ \sigma_{j}^{z} + (\sigma_{j}^{+}\! +\! 
                 \sigma_{j}^{-})\cdot
                 (\sigma_{j+1}^{+}\! +\! \sigma_{j+1}^{-}) \right] \;
                 + \sigma_{N}^{z} + \alpha \cdot (\sigma_{N}^{+}\! +\! 
                 \sigma_{N}^{-})\cdot
                 (e^{i\phi}\sigma_{1}^{+}\! +\! e^{-i\phi}\sigma_{1}^{-}) 
                 \right) \; ,
\label{eq:H}
\end{equation}
where $N$ denotes the number of sites, $\sigma_{j}^{\pm} = \frac{1}{2}
(\sigma_{j}^{x}\! \pm \! i\sigma_{j}^{y})$, and $\alpha \! \in \! {\bf
R},\ \phi \! \in \! [0,\pi)$ .  This corresponds to taking in
Eq.~(\ref{I:gd}) \mbox{$e_{1} \! = \! e_{z} \! = \! 0$} and
\mbox{$e_{x} \! = \! \alpha \cos (\phi ) \; , \; e_{y} \! = \! -\alpha
\sin (\phi )$}.  The Hamiltonian $H(\alpha,\phi)$ (\ref{eq:H}) for all
values of the parameters commutes with the charge operator $\hat{Q}$
(\ref{e:Q}) defined in Sec.~2.  In order to avoid the appearance of a
nonlocal number operator and in complete analogy to the known case
$\phi \! =\! 0$ \cite{hps} (see also Sec.~2), one considers the mixed
sector Hamiltonian $\tilde{H}(\alpha ,\phi )$~(\ref{e:P})
\begin{equation}
\tilde{H}(\alpha ,\phi ) = 
H(\alpha ,\phi )\cdot P_{0} + H(-\alpha ,\phi )\cdot P_{1} . 
\label{eq:Ht}
\end{equation}
Here, the $P_{q}\;$, $\; q \! \in \! \{0,1\}$, are the projection
operators formerly defined in Eq.~(\ref{e:P}).  After performing a
Jordan-Wigner transformation \cite{lsm}
\begin{equation}
\begin{array}{ccccc}
c_{k} & = & \left( {\displaystyle \prod_{j=1}^{k-1}} 
\sigma_{j}^{z} \right) \sigma_{k}^{-} & = &
        \left( {\displaystyle \prod_{j=1}^{k-1}} 
\exp (i\pi \sigma_{j}^{-}\sigma_{j}^{+} ) \right) \sigma_{k}^{-}
\vspace{4mm} \\
c_{k}^{\dagger} & = & \left( {\displaystyle \prod_{j=1}^{k-1}} 
\sigma_{j}^{z} \right) \sigma_{k}^{+} & = &
        \left( {\displaystyle \prod_{j=1}^{k-1}} 
\exp (-i\pi \sigma_{j}^{-}\sigma_{j}^{+} ) \right) \sigma_{k}^{+}
\label{eq:cc}
\end{array}
\end{equation}
the Hamiltonian $\tilde{H}(\alpha ,\phi )$ (\ref{eq:Ht}) reads
\begin{equation}
\tilde{H}(\alpha ,\phi ) = \frac{N}{2} + \sum_{j,k=1}^{N} 
                   \left[ B_{j,k}(\alpha ,\phi )c_{j}^{\dagger}c_{k}
                   + \frac{1}{2} \left( B_{j,k}^{\prime}(\alpha ,\phi )
                     c_{j}^{\dagger}c_{k}^{\dagger}
                   + B_{k,j}^{{\prime}^{\scriptstyle\ast}}(\alpha,\phi )
                     c_{j}c_{k} \right) \right] \; \; ,
\label{eq:HT}
\end{equation}
where the two $N\! \times \! N$ matrices $B(\alpha ,\phi )$ and
$B^{\prime}(\alpha ,\phi )$ are given by
\begin{equation}
\begin{array}{lcl}
B_{j,k}(\alpha ,\phi ) & = &  -\delta_{j,k} + \frac{1}{2}\delta_{j+1,k} 
                              (1\! -\! \delta_{j,N} )
                              + \frac{1}{2}\delta_{j,k+1} 
                              (1\! -\! \delta_{k,N} ) \vspace{2mm} \\
 & &                          \mbox{} - \frac{\alpha}{2}
                              (e^{i\phi}\delta_{j,1}
                              \delta_{k,N}\! +\! e^{-i\phi}
                              \delta_{j,N}\delta_{k,1} )
\vspace{4mm} \\
B_{j,k}^{\prime}(\alpha ,\phi )
                       & = &  \frac{1}{2}\delta_{j+1,k} 
                              (1\! -\! \delta_{j,N} )
                             - \frac{1}{2}\delta_{j,k+1} 
                              (1\! -\! \delta_{k,N} )
                             + \frac{\alpha}{2}e^{i\phi} 
                              (\delta_{j,1}\delta_{k,N}\! -\! 
                               \delta_{j,N}\delta_{k,1} )
\label{eq:BB}
\end{array}
\end{equation}
and the fermionic operators $c_{k}$ and $c_{k}^{\dagger}$ fulfill the
anticommutation relations
\begin{equation}
\begin{array}{ccccc}
\{ c_{k},c_{k^{\prime}} \} & = & 
\{ c_{k}^{\dagger},c_{k^{\prime}}^{\dagger} \} & = & 0 \vspace{4mm} \\
\{ c_{k},c_{k}^{\prime^{\scriptstyle \dagger}} \} & = & 
\delta_{k,k^{\prime}} \; \; .  & &
\end{array} \label{eq:acr}
\end{equation}
The matrices $B(\alpha ,\phi )$ and $B^{\prime}(\alpha ,\phi )$ have
the properties
\begin{equation}
\begin{array}{ccccc}
B^{\dagger}(\alpha , \phi ) & = & B(\alpha , \phi ) & & \vspace{4mm} \\
B^{\prime ^{\scriptstyle \dagger}}(\alpha , \phi ) & = &
- B^{\prime ^{\scriptstyle \ast}}(\alpha , \phi ) & = & 
- B^{\prime}(\alpha , -\phi )
\end{array}
\end{equation}
and, in particular, are not real except for $\phi \!= \! 0$ or $\alpha
\! =\! 0$ which correspond to the previously examined case of one
(usual) defect \cite{hps} .
 
In the sequel the Hamiltonian $\tilde{H}(\alpha , \phi )$
(\ref{eq:HT}) which is bilinear in the fermionic operators $c_{k}$ and
$c_{k}^{\dagger}$ (\ref{eq:cc}) shall be diagonalized to the form
\begin{equation}
\tilde{H}(\alpha , \phi ) = \sum_{j=1}^{N} 
\Lambda_{j}(\alpha , \phi )a_{j}^{\dagger}a_{j}^{\mbox{}} +
E_{0}(\alpha,\phi,N)
\label{eq:HTD}
\end{equation}
by means of a Bogoliubov transformation, where the operators $a_{k}$
and $a_{k}^{\dagger}$ fulfill the same anticommutation relations as
$c_{k}$ and $c_{k}^{\dagger}$ (\ref{eq:acr}), respectively, and the
$\Lambda_{j}(\alpha , \phi )$ are real numbers for all $j\! =\! 1,
\ldots ,N$ due to the hermiticity of the Hamiltonian.
$E_{0}(\alpha,\phi,N)$ denotes the ground-state energy that depends on
the definition of the vacuum.  For this purpose we make the ansatz
\begin{equation}
\begin{array}{ccc}
a_{k}           & = & {\displaystyle \sum_{j=1}^{N}} 
( g_{k,j}c_{j} + h_{k,j}^{\ast}c_{j}^{\dagger} ) \vspace{4mm} \\
a_{k}^{\dagger} & = & {\displaystyle \sum_{j=1}^{N}} 
( g_{k,j}^{\ast}c_{j}^{\dagger} + h_{k,j}c_{j} )
\end{array}
\label{eq:bt}
\end{equation}
with $k\! =\! 1, \ldots ,N$ and (in general) complex coefficients
$g_{k,j},h_{k,j} \in {\bf C}$. This generalization compared to the
transformation used in \cite{lsm,hps} is enforced by the non-reality
of the Hamiltonian~(\ref{eq:HT}).  From the anticommutation relations
for the operators $a_{k}$ and $a_{k}^{\dagger}$ (\ref{eq:acr}), one
yields conditions on these coefficients which are in fact equivalent
to the $2N\! \times \! 2N$ matrix $U$ defined by
\begin{equation}
U = \left(  \begin{array}{cc}
                g & h^{\ast} \\
                h & g^{\ast}
            \end{array}                    \right)
\label{eq:U}
\end{equation}
being unitary. Here, $g$ and $h$ denote $N\! \times \! N$ matrices
with the entries $g_{k,j}$ and $h_{k,j}$ ($k,j\! =\! 1, \ldots ,N$),
respectively. In other words, one has to find a unitary transformation
$U$ that diagonalizes the Hamiltonian~(\ref{eq:HT}) with the
additional property
\begin{equation}
C U C = U^{\ast} \; \; ,
\label{eq:C}
\end{equation}
$C$ denoting the $2N\! \times \! 2N$ matrix
\begin{equation}
\begin{array}{ccc}
C = \left( \begin{array}{cc} {\bf 0}_{N} & {\bf 1}_{N} \\ {\bf 1}_{N}
& {\bf 0}_{N} \end{array} \right) & , & C^{2} = {\bf 1}_{2N} \; \; .
\end{array}
\label{eq:CC}
\end{equation}
Note that as a consequence of this property the determinant of $U$ is
restricted to the values \mbox{$\det U \in \{1,-1\}$}.
 
Necessary conditions for the existence of a transformation as desired
are the $N$ equations
\begin{equation}
\begin{array}{ccc}
[a_{k},\tilde{H}(\alpha , \phi )] = \Lambda_{k}(\alpha , \phi ) 
\cdot a_{k} & , & k\! =\! 1, \ldots ,N \; ,
\end{array}
\label{eq:aH}
\end{equation}
which, using Eqs.~(\ref{eq:HT})~and~(\ref{eq:bt}), after some
calculations result in the equations \mbox{($k\! =\! 1, \ldots ,N$)}
\begin{equation}
\begin{array}{lcc}
\Lambda_{k} \cdot \vec{g}_{k}        & = & 
\vec{g}_{k} \cdot B - \vec{h}_{k}^{\ast}\cdot 
B^{\prime^{\scriptstyle \ast}}
\vspace{4mm} \\
\Lambda_{k} \cdot \vec{h}_{k}^{\ast} & = & 
\vec{g}_{k} \cdot B^{\prime} - \vec{h}_{k}^{\ast}\cdot B^{\ast}
\end{array}
\label{eq:Lgh}
\end{equation}
together with their complex conjugated. For convenience, the
dependency on the parameters $\alpha$ and $\phi$ was dropped and the
row vectors \mbox{$\vec{g}_{k}=(g_{k,1}, \ldots ,g_{k,N})$} and
\mbox{$\vec{h}_{k}=(h_{k,1}, \ldots ,h_{k,N})$} were used. Introducing
the notations
\begin{equation}
\begin{array}{ccc}
\vec{\Phi}_{k} & = & \frac{1}{\sqrt{2}} \left( g_{k,1}\! +\! h_{k,1}, 
\ldots ,g_{k,N}\! \! +h_{k,N},
g_{k,1}^{\ast}\! +\! h_{k,1}^{\ast}, \ldots ,
g_{k,N}^{\ast}\! +\! h_{k,N}^{\ast} \right) \vspace{4mm} \\
\vec{\Psi}_{k} & = & \frac{i}{\sqrt{2}} \left( g_{k,1}\! -\! h_{k,1}, 
\ldots ,g_{k,N}\! -\! h_{k,N},
h_{k,1}^{\ast}\! -\! g_{k,1}^{\ast}, \ldots ,
h_{k,N}^{\ast}\! -\! g_{k,N}^{\ast} \right) \\
\end{array}
\label{eq:pp}
\end{equation}
and
\begin{equation}
\widehat{B} = \left( \begin{array}{cc} B & B^{\prime} \\ 
-B^{\prime^{\scs \ast}} & -B^{\ast} \end{array} \right) \; \; ,
\label{eq:BBT}
\end{equation}
with $\widehat{B}$ being hermitian, Eqs.~(\ref{eq:Lgh}) become
\begin{equation}
\begin{array}{ccrc}
\Lambda_{k} \cdot \vec{\Phi}_{k} & = &
 -i \vec{\Psi}_{k} \cdot \widehat{B} \vspace{4mm} & \\
\Lambda_{k} \cdot \vec{\Psi}_{k} & = & 
 i \vec{\Phi}_{k} \cdot \widehat{B} & .
\end{array}
\label{eq:LL2}
\end{equation}
One therefore derives the following eigenvalue equations for the
squares of the fermion frequencies $\Lambda_{k}$ \mbox{($k\! =\!
1,\ldots ,N$)}
\begin{equation} 
\begin{array}{ccccl}
\Lambda_{k}^{2} \cdot \vec{\Phi}_{k} & = & 
\vec{\Phi}_{k} \cdot \widehat{A} 
& = & \vec{\Phi}_{k} \cdot \widehat{B}^{2} \vspace{4mm} \\
\Lambda_{k}^{2} \cdot \vec{\Psi}_{k} & = & 
\vec{\Psi}_{k} \cdot \widehat{A} 
& = & \vec{\Psi}_{k} \cdot \widehat{B}^{2} \; \; ,
\label{eq:L2} \end{array}
\end{equation}
where $\widehat{A}\! =\! \widehat{B}^{2}$ is an --- at least ---
positive semi-definite hermitian $2N\! \times \! 2N$ matrix given
explicitly by
\begin{equation}
\widehat{A}(\alpha , \phi ) = 
\left( \begin{array}{cc} A(\alpha , \phi ) & 
A^{\prime}(\alpha , \phi) \\
A^{\prime^{\scs \ast}}(\alpha , \phi ) & 
A^{\ast}(\alpha , \phi ) \end{array} \right)
\label{eq:AT}
\end{equation}
with the two $N\! \times \! N$ matrices $A$ and $A^{\prime}$
\begin{equation}
\begin{array}{ccl}
A_{j,k}(\alpha , \phi ) & = &
     \delta_{j,k} \cdot
     \left( 2-\frac{1-\alpha^{2}}{2}(\delta_{j,1}\delta_{k,1}\! +\! 
     \delta_{j,N}\delta_{k,N}) \right)
     -\delta_{j+1,k} \cdot (1\! -\! \delta_{j,N})
     -\delta_{j,k+1} \cdot (1\! -\! \delta_{k,N}) \vspace{2mm} \\
 & & \mbox{} + \alpha e^{i\phi}\delta_{j,1}\delta_{k,N}
     +\alpha e^{-i\phi}\delta_{j,N}\delta_{k,1}
     -\frac{i\alpha \sin (\phi )}{2} \cdot (\delta_{j,2}
     \delta_{k,N}\! -\! \delta_{j,N}\delta_{k,2})
\vspace{5mm} \\
A_{j,k}^{\prime}(\alpha , \phi ) & = &
     \frac{\alpha^{2}e^{2i\phi}-1}{2}\delta_{j,1}\delta_{k,1}
     -\frac{\alpha^{2}-1}{2}\delta_{j,N}\delta_{k,N}
     +\frac{i\alpha \sin (\phi )}{2} \cdot 
     (\delta_{j,2}\delta_{k,N}\! +\! \delta_{j,N}\delta_{k,2}) \; \; .
\end{array}
\label{eq:AA}
\end{equation}
Note the following property of $\widehat{A}(\alpha ,\phi )$
\begin{equation}
C \widehat{A}(\alpha ,\phi ) C = 
\widehat{A}^{\ast}(\alpha  ,\phi ) \; \; .
\label{eq:CAC}
\end{equation}
 
The unitarity condition on $U$ (\ref{eq:U}) leads to the restrictions
\begin{equation}
\begin{array}{ccc}
\vec{\Phi}_{j} \cdot \vec{\Phi}_{k}^{\dagger} =
\vec{\Psi}_{j} \cdot \vec{\Psi}_{k}^{\dagger} = \delta_{j,k} & , &
\vec{\Phi}_{j} \cdot \vec{\Psi}_{k}^{\dagger} = 0 \; \; ,
\end{array}
\label{eq:PP}
\end{equation}
or, equivalently, to the unitarity of the matrix $\widehat{U}$,
\begin{equation}
\widehat{U} = \left( \begin{array}{cc} g+h & (g+h)^{\ast} \\ 
i(g-h) & (i(g-h))^{\ast} \end{array} \right) = T \cdot U \; \; ,
\label{eq:UT}
\end{equation}
that is built from the row vectors $\vec{\Phi}_{k}$ and
\mbox{$\vec{\Psi}_{k} , \; k\! =\! 1 ,\ldots , N$}. Here, $T$ is the
unitary matrix
\begin{equation}
T = \frac{1}{\sqrt{2}} \left( 
\begin{array}{cc} {\bf 1}_{N} & {\bf 1}_{N} \\ 
i{\bf 1}_{N} & -i{\bf 1}_{N} \end{array} \right) \; \; .
\label{eq:T}
\end{equation}
In addition, the fact that $U$ has to comply with Eq.~(\ref{eq:C})
leads to the conditions
\begin{equation}
\begin{array}{ccccc}
\vec{\Phi}_{k} \cdot C = \vec{\Phi}_{k}^{\ast} & , &
\vec{\Psi}_{k} \cdot C = \vec{\Psi}_{k}^{\ast} & , & 
k\! =\! 1, \ldots , N \; \; ,
\end{array}
\label{eq:C2}
\end{equation}
or, equivalently,
\begin{equation}
\widehat{U} \cdot C = \widehat{U}^{\ast} \; \; .
\label{eq:UC}
\end{equation}
However, in Appendix~A it is shown that the latter conditions can be
fulfilled by a simple change of basis always.  Thus it is possible to
reconstruct from the knowledge of $\widehat{U}$ the transformation $U$
(\ref{eq:U}) by $U = T^{\dagger} \widehat{U}$ and thereby the
Eqs.~(\ref{eq:LL2}), (\ref{eq:PP}), and (\ref{eq:C2}) in fact are
necessary and sufficient to determine the Bogoliubov
transformation~(\ref{eq:bt}).
 
In this way the determination of the $\Lambda_{k}$ --- up to a sign
--- has been transformed to the problem of computing the eigenvalues
of a $2N\! \times \! 2N$ hermitian matrix and, furthermore, a
Bogoliubov transformation~(\ref{eq:bt}) can be constructed from the
knowledge of the eigenvectors. The only thing that has been missed so
far is the degeneration of the spectrum of $\widehat{A}$ that has to
exist if the Eqs.~(\ref{eq:L2}) can be solved consistently. This is
the purpose of Appendix~B, where the origin of this property ---
obviously a symmetry hidden in the equations --- is examined. One
might think that this is not necessary because it follows from the
ansatz~(\ref{eq:bt}) , but it is the aim to show that this ansatz can
always be used and does not lead to equations that cannot be solved.
The question about the sign of the $\Lambda_{k}$ can be answered as
follows. A change of the sign of $\Lambda_{k}$ corresponds to an
exchange of the corresponding creation and annihilation operators
$a_{k}^{\dagger}$ and $a_{k}$, which is a special Bogoliubov
transformation (see also Appendix~B), together with a shift in the
ground-state energy $E_{0}$ (\ref{eq:HTD}).  In order to let the
vacuum (the state annihilated by all annihilation operators) be the
state with lowest energy, one chooses all $\Lambda_{k}$ to be
non-negative.
 
Let me stop at this point for a short comment.  From
Eq.~(\ref{eq:AA}), one recognizes that the matrix $A^{\prime}(\alpha ,
\phi )$ does not vanish for $\phi \! =\! 0$ as one might have
expected.  This is of course due to the choice of the basis
(\ref{eq:pp}) and it is easy to find a Bogoliubov transformation
(\ref{eq:bt}) that block-diagonalizes $\widehat{A}(\alpha ,0)$
(\ref{eq:AT}) to a form with twice the same real symmetric matrix
along the diagonal.  In this way the problem as usually reduces to the
diagonalization of a $N\! \times \! N$ matrix \cite{hps} and one just
obtains the whole spectrum doubled.  Principally, the
block-diagonalization remains possible in the general case $\phi \!
\neq \! 0$, too, but the transformation may become arbitrarily
complicated. However, in Appendix~B it will be argued that the
doubling of the spectrum is a general feature --- as it must be to
fulfill Eqs.~(\ref{eq:L2}) --- not limited to the case $\phi \! = \!
0$.
 
The eigenvalue equations~(\ref{eq:L2}) are now solved by means of an
ansatz. This is done the same way as in \cite{lsm,hps,schl} by
\begin{equation} 
\begin{array}{lcl}
\vec{\Phi}_{p,j} & = & \frac{1}{\sqrt{2}} \cdot
\left( r_{j} e^{\frac{ip}{N}} + 
s_{j} e^{-\frac{ip}{N}} \right) \vspace{4mm} \\
\vec{\Phi}_{p,j+N} & = & \frac{1}{\sqrt{2}} \cdot
\left( \tilde{r}_{j} e^{\frac{ip}{N}} + 
\tilde{s}_{j} e^{-\frac{ip}{N}} \right)
\end{array}
\label{eq:ans}
\end{equation}
with \mbox{$j\! = \! 1,\ldots ,N$}, where the eigenvectors
$\vec{\Phi}_{p}$ (and likewise the eigenvalues $\Lambda_{p}$) are from
now on indexed by the variable $p$ which will (in general) not be
integer-valued.  One observes that all but six of the $2N$ linear
homogeneous equations summarized in
\begin{equation}
\Lambda_{p}^{2} \cdot \vec{\Phi}_{p} = 
\vec{\Phi}_{p} \cdot \widehat{A}(\alpha ,\phi )
\label{eq:L3}
\end{equation}
are solved by the relation
\begin{equation}
\Lambda_{p}^{2} = \left( 2 \sin (\frac{p}{2}) \right) ^{2} \; \; .
\label{eq:Lp}
\end{equation}
The remaining six equations then determine the possible values of $p$
through the requirement that there has to exist a solution for the
variables $r_{j}$, $\tilde{r}_{j}$, $s_{j}$, and $\tilde{s}_{j}$ in
the ansatz~(\ref{eq:ans}). This is obviously equivalent to demand the
coefficient determinant to vanish. The computation of this determinant
yields a --- highly non-linear --- equation for the allowed values of
$p$ and therewith for the eigenvalues $\Lambda_{p} \; $.  It is known
that in this way one obtains all those eigenvalues $\Lambda_{p}$ whose
squares are smaller than four \cite{lsm}.  However, this is sufficient
to determine the finite-size scaling limit, because in that case only
low-energy states enter the calculation.
 
Of course, one is not able to solve the equation that determines $p$
for any size $N$ of the chain. One is rather interested in the
thermodynamic limit \mbox{$N \rightarrow \infty$}.  This is performed
by expanding the determinant in inverse powers of $N$, where one uses
\cite{hps,bcs,schl}
\begin{equation}
N \cdot p = \varepsilon_{p} + O(N^{-1}) \; \; .
\label{eq:pN1}
\end{equation}
The coefficient of the first non-vanishing power in $N^{-1}$ leads to
an equation for $\varepsilon_{p}$ that determines the scaled
eigenvalues in the continuum limit by
\begin{equation}
\lim_{N \rightarrow \infty} \left( N \cdot \Lambda_{p} \right) =
\varepsilon_{p} \; \; .
\label{eq:sl}
\end{equation}
 
The above-mentioned algebraic calculations were performed on the
computer. Let me only state the result.  The final equation for
\mbox{$\varepsilon_{p}(\alpha ,\phi )$} reads
\begin{equation}
\left( (1 + \kappa^{2}) \cdot 
\cos (\varepsilon_{p}) + 2 \kappa \right) ^{2} = 0 \; \; \; ,
\; \; \; \kappa = \alpha \cos(\phi) \; \; ,
\label{eq:kap}
\end{equation}
and one recognizes that this is just the square of the equation that
one obtains for one `ordinary' defect of strength \mbox{$\kappa \! =
\! \alpha \cos (\phi )$} \cite{hps,bcs}.  Therefore the finite-size
scaling limit is the same for both systems, although they generally
differ on finite chains. Thus the operator content that only depends
on $\kappa$ can be read off from Eqs.~(\ref{e:ek}) and (\ref{e:phi})
in Sec.~2 directly.
 
Therefore, exactly as for an `ordinary' defect \cite{bcs,schl}, the
scaled (relative to the ground-state energy $E_{0}(1,0,N)$ for
periodic BCs) Hamiltonian \mbox{$\widehat{\tilde{H}}(\alpha,\phi)$}
\begin{equation}
\widehat{\tilde{H}}(\alpha,\phi) = 
\lim_{N \rightarrow \infty} \frac{N}{2\pi} \cdot
\left( \tilde{H}(\alpha,\phi) - E_{0}(1,0,N) + A_{1}(\alpha,\phi) \right)
\label{eq:hh} 
\end{equation}
is given by a generator $\tilde{L}_{0}(\varphi(\alpha \cos(\phi)))$ of
a shifted $u(1)$ KMA (\ref{e:shift}).  From the positive solutions of
Eq.~(\ref{eq:kap}) one obtains the scaled values for the fermion
frequencies (\ref{eq:sl}) which have to be inserted into
Eq.~(\ref{eq:HTD}). This results in
\begin{equation} 
\begin{array}{ccl}
\widehat{\tilde{H}}(\alpha,\phi)
& = & \widehat{\tilde{H}}(\kappa) \vspace{4mm} \\
& = & \dys{\sum_{k \in {\bf N}_{0}}} 
\left( k+\frac{1}{2} - \varphi(\kappa) \right) a_{k}^{\dagger} a_{k}
+ \left( k+\frac{1}{2} + \varphi(\kappa) \right) b_{k}^{\dagger} b_{k} \; 
+ \frac{1-2\tilde{c}(\kappa)}{24} \vspace{4mm} \\
& = & \tilde{L}_{0}(\varphi(\kappa))
\; \; ,
\end{array} \label{eq:hd} 
\end{equation}
where the shift $\varphi(\kappa)$ has the same form as for an
`ordinary' defect \cite{hps,bcs} (cf. Eq.~(\ref{e:phi}))
\begin{equation}
\varphi(\kappa) = \frac{1}{4} - \frac{1}{\pi} \arctan (\kappa)  \; \; ,
\label{eq:phi} 
\end{equation}
with \mbox{$\kappa \! = \! \alpha \cos(\phi)\;$}.  Here,
Eq. (\ref{e:gse}) with \mbox{$A_{2}(1)\!=\! \pi c/6$} and the
expansion of the ground-state energy $E_{0}(\alpha,\phi,N)$
(\ref{eq:HTD}) of the Hamiltonian $\tilde{H}(\alpha,\phi)$
(\ref{eq:HT})
\begin{equation}
-E_{0}(\alpha,\phi,N) = A_{0} \cdot N + A_{1}(\alpha,\phi) + 
\frac{\pi \tilde{c}(\alpha,\phi)}{6} \cdot N^{-1} + o(N^{-1})
\label{eq:gse}
\end{equation}
were used. The `effective' central charge $\tilde{c}(\alpha,\phi)$
that also depends on \mbox{$\kappa\!=\!\alpha \cos(\phi)$} only, is
determined by the shift $\varphi$ (\ref{eq:phi}) through the relation
\begin{equation}
\frac{\left( \varphi(\kappa) \right) ^{2}}{2} = 
\frac{1-2\tilde{c}(\kappa)}{24} \; \; ,
\label{eq:ct}
\end{equation}
and the fermionic operators $a_{k}$ and $b_{k}$ in Eq.~(\ref{eq:hd})
descend from the $a_{k}$ in Eq.~(\ref{eq:HTD}) by means of a suitable
re-numeration of the indices.
 
The only thing involved in the resulting expressions that does not
depend on \mbox{$\kappa\! = \! \alpha \cos (\phi)$} alone is the
(non-universal) surface energy $A_{1}(\alpha,\phi)$. Since the
ground-state energy does not obtain from the analysis performed, no
analytic expression is known for this term.  However, numerical
investigations show that it differs from the $A_{1}(\alpha
\cos(\phi))$ of Eq.~(\ref{e:A1}) by an amount of $\frac{1}{\pi}$ at
most (for \mbox{$\phi \! = \frac{\pi}{2}$} and \mbox{$\alpha
\rightarrow \infty$}), thus it resembles the function for an
`ordinary' defect. In analogy to the case of `ordinary' defects, a
generalization of this investigation to an arbitrary number of
isolated generalized defects of this form at commensurate distances
\cite{hps,bcs} or to a generalized extended defect \cite{haye}
obviously can be performed the same way as for `ordinary' defects.

\section{Finite-size non-integrable generalized defects}
\setcounter{equation}{0}
 
In this section another two-parameter family of Hamiltonians
\mbox{${\cal H}(\beta ,\psi)$} defined by
\begin{equation}
{\cal H}(\beta ,\psi ) = 
-\frac{1}{2} \left( \sum_{j=1}^{N-1} 
\left[ \sigma_{j}^{z} + \sigma_{j}^{x}\sigma_{j+1}^{x} \right]
\; + \sigma_{N}^{z} + \beta \sigma_{N}^{x} \cdot
\left( \cos(\psi) \sigma_{1}^{x} - \sin(\psi) \sigma_{1}^{z} \right)
 \right) \; \; ,
\label{:H} 
\end{equation}
will be considered. Here, \mbox{$\beta \! \in \! {\bf R}\! - \! \{ 0
\}$} and \mbox{$\psi \! \in \! (0,\pi )$}, whereby one guarantees that
\mbox{$\beta \sin (\psi)$} is different from zero because \mbox{$\beta
\sin (\psi) \! = \! 0$} recovers finite-size integrable defects
examined in the preceding section.  As has been discussed in Sec.~2,
the Hamiltonian~(\ref{:H}) is not finite-size integrable under this
circumstances.  This is caused by the occurrence of the term
\mbox{$\sigma_{N}^{x} \sigma_{1}^{z}$} that not only breaks the global
${\bf Z}_2$ invariance of the Ising chain Hamiltonian (see
Eq.~(\ref{e:Q})), but also leads to the appearance of a three-fermion
coupling if a Jordan-Wigner transformation as in Eq.~(\ref{eq:cc}) is
performed on ${\cal H}$~(\ref{:H}).
 
Since for this reason the methods used in the previous section can not
be applied, the Hamiltonian~(\ref{:H}) has been investigated by
numerical calculations only.  This was done by diagonalizing ${\cal
H}$ for up to sixteen sites using an algorithm by Lanczos
\cite{la}. The scaled energy gaps $\overline{F}_{k}(\beta, \psi,N)$,
defined in this case relative to free BCs
\begin{equation}
\overline{F}_{k}(\beta ,\psi ,N) = \frac{N}{\pi}
\left( E_{k}(\beta ,\psi ,N) + 
A_{1}(\beta , \psi) - E_{0}(0,0,N) - A_{1}(0,0) \right)
\label{:F} 
\end{equation}
with \mbox{$A_{1}(0,0) \! = \! \frac{1}{\pi}\! - \! \frac{1}{2}$}
(\ref{e:A1}), have to be extrapolated to an infinite system
\begin{equation}
\overline{\cal F}_{k}(\beta ,\psi) = 
\lim_{N \rightarrow \infty} \overline{F}_{k}(\beta ,\psi ,N) \; \; .
\label{:Fc} 
\end{equation}
Here, $A_{1}(\beta,\psi)$ denotes the surface term that occurs in the
expansion of the lowest eigenvalue $E_{0}(\beta,\psi,N)$
(cf. Eq.~(\ref{e:gse})) of the Hamiltonian~(\ref{:H})
\begin{equation}
- E_{0}(\beta,\psi,N) = A_{0} \cdot N + 
A_{1}(\beta,\psi) + \frac{\pi \tilde{c}(\beta,\psi)}{24} \cdot N^{-1}
                        + o(N^{-1}) \;\; ,
\label{:c} 
\end{equation}
and $E_{k}(\beta,\psi,N)$ is the $k^{th}$ eigenvalue of the $N$-site
Hamiltonian ${\cal H}(\beta,\psi)$~(\ref{:H}).  The partition function
\mbox{${\cal F}(\beta,\psi,z)$} for the finite-size scaling spectrum
is defined by
\begin{equation}
{\cal F}(\beta,\psi,z) = \lim_{N \rightarrow \infty} 
\sum_{k} z^{\overline{F}_{k}(\beta,\psi,N)} \; \; ,
\label{:ss} 
\end{equation}
where the sum extends over the whole spectrum of ${\cal H}$
(\ref{:H}).  The use of $E_{0}(0,0,N)$ as the reference energy in
Eq.~(\ref{:F}) will be justified by the final expression for the
partition function \mbox{${\cal F}(\beta,\psi,z)$}~(\ref{:ss})
conjectured on the basis of numerical results.
 
There is one disadvantage in the use of Eqs.~(\ref{:F}) and
(\ref{:Fc}) for the actual numerical computations, namely that the
analytic form of the surface energy $A_{1}(\beta,\psi)$ (\ref{:c}) is
not known. Therefore, one alternatively extrapolates the scaled energy
gaps referred to the relative ground-state $E_{0}(\beta,\psi,N)$ and
besides that determines the lowest scaled energy gap
\mbox{$\overline{\cal F}_{0}(\beta,\psi)$}
\begin{equation}
\overline{\cal F}_{0}(\beta,\psi) = 
\lim_{N \rightarrow \infty} \frac{N}{\pi} \cdot
\left( E_{0}(\beta ,\psi ,N) - E_{0}(0,0,N)  + 
A_{1}(\beta , \psi) - A_{1}(0,0) \right)
\label{:F0} 
\end{equation}
from the finite-size corrections of the relative ground-state energy
$E_{0}(\beta,\psi,N)$~(\ref{:c}).  Knowing the `effective central
charge' $\tilde{c}(\beta,\psi)$ (\ref{:c}) is equivalent to the
knowledge of \mbox{$\overline{\cal F}_{0}(\beta,\psi)$}, because by
Eqs.~(\ref{e:gse}) \mbox{(with $A_{2}(0)\! = \! \frac{\pi c}{24}$)}
and (\ref{:F0}) the relation \mbox{($c\! =\! \frac{1}{2}$)}
\begin{equation} \begin{array}{ccc}
\overline{\cal F}_{0}(\beta,\psi) & = & 
       \frac{c - \tilde{c}(\beta,\psi)}{24} \vspace{2mm} \\
 & = & \frac{1 - 2 \tilde{c}(\beta,\psi)}{48}
\end{array} \label{:x}
\end{equation}
holds. This was used to numerically compute \mbox{$\overline{\cal
F}_{0}(\beta,\psi)$} for various choices of the parameters.  For this
purpose the known value of \mbox{$A_{0}\! = \! \frac{2}{\pi}$} was
inserted into Eq.~(\ref{:c}) and the unknown surface term
\mbox{$A_{1}(\beta,\psi)$}~(\ref{:c}) was eliminated by combining the
equation for two subsequent values of $N$.  The value of
\mbox{$\overline{\cal F}_{0}(\beta,\psi)$} then could be determined
approximatively by extrapolating to infinite size \mbox{$N \rightarrow
\infty$}, wherefore the algorithm developed by Bulirsch and Stoer
(\cite{bs}, see also \cite{hs}) was employed.
 
The numerical results are in reasonable agreement with the following
conjecture for the operator content expressed through the character
functions \cite{rc} $\chi_{\Delta}(z)$ (\ref{e:cf}) of the unitary
irreps of the \mbox{$c\!  = \! \frac{1}{2}$} VA~(\ref{e:VA})
corresponding to a highest weight $\Delta$ , \mbox{$\, \Delta \! \in
\! \{0,\frac{1}{16},\frac{1}{2}\}$}.  For \mbox{$\beta \! > \! 0$} one
has
\begin{equation}
{\cal F}(\beta,\psi,z) \; \;  = \; \;  
\left\{ \begin{array}{ll}
\chi_{0}(z)            & 
\mbox{ if $\; \psi \! \in \! (0,\frac{\pi}{2})$} \vspace{2mm} \\
\chi_{\frac{1}{16}}(z) & 
\mbox{ if $\; \psi \! = \! \frac{\pi}{2}$} \vspace{2mm} \\
\chi_{\frac{1}{2}}(z)  & 
\mbox{ if $\; \psi \! \in \! (\frac{\pi}{2},\pi)$} \end{array} \right.
\label{:oc} 
\end{equation}
wherefrom the case \mbox{$\beta \! < \! 0$} can be deduced using the
relation
\begin{equation}
{\cal F}(\beta,\psi,z)  = {\cal F}(-\beta,\pi-\psi,z) \; \; .
\label{:bn} 
\end{equation}
This equality follows directly from the fact that for arbitrary $N$
the Hamiltonian~(\ref{:H}) possesses the property
\begin{equation}
\hat{Q} \cdot {\cal H}(\beta,\psi) \cdot \hat{Q} = {\cal H}(-\beta,\pi-\psi)
\label{:QHQ} 
\end{equation}
implying the equality of the finite-size spectra of \mbox{${\cal
H}(\beta,\psi)$} and \mbox{${\cal H}(-\beta,\pi-\psi)$}.
 
\begin{table}[htb]
\caption{Opposition of numerical estimates and by
Eq.~(\ref{:oc}) conjectured values of the 
lowest scaled energy gap
\mbox{$\overline{\cal F}_{0}(\beta,\psi)$} 
(\ref{:F0}) for \mbox{$\beta \! = \! 1$} 
and different choices of the parameter 
$\psi$.}\vspace{1ex}
\centering
\begin{tabular}{|c|l|c|} 
\hline 
  $\psi$  & numerical estimate           & conjectured value \\ 
\hline
$0.1 \pi$ & \hspace*{7.5mm} $0.0005(3)$  &   $0.000000$ \\
$0.2 \pi$ & \hspace*{7.5mm} $0.0001(2)$  &   $0.000000$ \\
$0.3 \pi$ & \hspace*{7.5mm} $0.0002(3)$  &   $0.000000$ \\
$0.4 \pi$ & \hspace*{7.5mm} $0.001(2)$   &   $0.000000$ \\ 
$0.5 \pi$ & \hspace*{7.5mm} $0.06250(3)$ &   $0.062500$ \\
$0.6 \pi$ & \hspace*{7.5mm} $0.6(1)$     &   $0.500000$ \\
$0.7 \pi$ & \hspace*{7.5mm} $0.498(1)$   &   $0.500000$ \\
$0.8 \pi$ & \hspace*{7.5mm} $0.4995(3)$  &   $0.500000$ \\
$0.9 \pi$ & \hspace*{7.5mm} $0.5(1)$     &   $0.500000$ \\
\hline
\end{tabular}
\end{table}

\begin{table}[htb]
\caption{The same as in Table~1, but for
three fixed values of the parameter $\psi$ and
varying  $\beta$.}\vspace{1ex}
\centering
\begin{tabular}{||r|c|l|c||} 
\hline \hline
$\beta$ \hspace*{2mm}
        & $\psi$   & numerical estimate           & conjectured value \\ 
\hline \hline
$0.5$   & $0.2\pi$ & \hspace*{7.5mm} $0.0002(3)$  &   $0.000000$      \\ 
$2.0$   & $0.2\pi$ & \hspace*{7.5mm} $0.00003(2)$ &   $0.000000$      \\
$5.0$   & $0.2\pi$ & \hspace*{7.5mm} $0.00001(2)$ &   $0.000000$      \\
$20.0$  & $0.2\pi$ & \hspace*{7.5mm} $0.00000(1)$ &   $0.000000$      \\
$100.0$ & $0.2\pi$ & \hspace*{7.5mm} $0.00000(1)$ &   $0.000000$      \\ 
\hline
$0.5$   & $0.5\pi$ & \hspace*{7.5mm} $0.06250(1)$ &   $0.062500$      \\
$2.0$   & $0.5\pi$ & \hspace*{7.5mm} $0.06251(2)$ &   $0.062500$      \\
$5.0$   & $0.5\pi$ & \hspace*{7.5mm} $0.06250(1)$ &   $0.062500$      \\
$20.0$  & $0.5\pi$ & \hspace*{7.5mm} $0.06250(1)$ &   $0.062500$      \\
$100.0$ & $0.5\pi$ & \hspace*{7.5mm} $0.0625(1)$  &   $0.062500$      \\ 
\hline
$0.5$   & $0.8\pi$ & \hspace*{7.5mm} $0.5(1)$     &   $0.500000$      \\
$2.0$   & $0.8\pi$ & \hspace*{7.5mm} $0.5005(3)$  &   $0.500000$      \\
$5.0$   & $0.8\pi$ & \hspace*{7.5mm} $0.50001(2)$ &   $0.500000$      \\
$20.0$  & $0.8\pi$ & \hspace*{7.5mm} $0.50000(1)$ &   $0.500000$      \\
$100.0$ & $0.8\pi$ & \hspace*{7.5mm} $0.50003(2)$ &   $0.500000$      \\
\hline \hline
\end{tabular}
\end{table}

In Table~1 the obtained estimates of the lowest scaled energy gap
\mbox{$\overline{\cal F}_{0}(\beta,\psi)$} (\ref{:x}) together with
their --- rather subjective --- errors for \mbox{$\beta \! = \! 1$}
and different values of $\psi$ are presented. Table~2 contains the
same thing for three fixed values of $\psi$, namely \mbox{$\pi/5 \; ,
\; \pi/2 \; $}, and $4\pi/5\;$, and varying $\beta$.  Of course, only
a small fraction of the collected data could be displayed therein.  In
both tables, the values deduced from the conjecture~(\ref{:oc}) are
opposed to the numerical data. A comparison shows that the agreement
is good on the whole, especially if one stays far enough away from
\mbox{$\beta \! = \! 0$} and \mbox{$\psi \! \in \! \{ 0 ,
\frac{\pi}{2} , \pi \}$}. Near those values of the parameters the
extrapolation is difficult (and the resulting `errors' thereby big),
what can be understood from the assertion that the dependence on the
parameters ceases to be continuous there.  Obviously one is not able
to see such an effect in a finite sample, thus the observed behaviour
agrees with what one would expect. Anyhow one might have awaited
possible discontinuities at those points where the symmetry of the
Hamiltonian (\ref{:H}) changes, these are all the critical values
mentioned above with the exception of \mbox{$\psi \! = \!
\frac{\pi}{2}$}. In this case no obvious change of the symmetry
occurs, but there might be a `hidden' symmetry that one cannot find by
inspection.  In Table~3 and Table~4, numerical results for the surface
energy $A_{1}(\beta,\psi)$~(\ref{:c}) for various values of the
parameters $\beta$ and $\psi$ are given. In this case no analytic
expression is known and thus there is nothing to compare with.
 
\begin{table}[htb]
\caption{Numerical estimates of the surface energy
$A_{1}(\beta,\psi)$ (\ref{:c}) for \mbox{$\beta \! = \! 1$}
and different values of $\psi$.}\vspace{1ex}
\centering
\begin{tabular}{|c|r@{.}l|} 
\hline 
  $\psi$  & \multicolumn{2}{c|}{numerical estimate} \\ 
\hline
$0.1 \pi$ & \hspace*{6.5mm} $0$ & $04377(2)$  \\
$0.2 \pi$ & \hspace*{6.5mm} $0$ & $07821(1)$  \\
$0.3 \pi$ & \hspace*{6.5mm} $0$ & $09682(3)$  \\
$0.4 \pi$ & \hspace*{6.5mm} $0$ & $10143(3)$  \\
$0.5 \pi$ & \hspace*{6.5mm} $0$ & $097454(3)$ \\
$0.6 \pi$ & \hspace*{6.5mm} $0$ & $101(1)$    \\
$0.7 \pi$ & \hspace*{6.5mm} $0$ & $095(2)$    \\
$0.8 \pi$ & \hspace*{6.5mm} $0$ & $080(1)$    \\
$0.9 \pi$ & \hspace*{6.5mm} $0$ & $0434(1)$   \\
\hline
\end{tabular}
\end{table}

\begin{table}[htb]
\caption{The same as in Table~3, but for three values
of the parameter $\psi$ and varying $\beta$.}\vspace{1ex}
\centering
\begin{tabular}{||r|c|r@{.}l||} 
\hline \hline
$\beta$ \hspace*{2mm}
        & $\psi$   & \multicolumn{2}{c||}{numerical estimate}  \\ 
\hline \hline
$0.5$   & $0.2\pi$ & \hspace*{3mm} $-0$ & $09663(3)$   \\ 
$2.0$   & $0.2\pi$ & \hspace*{3mm}  $0$ & $5105577(2)$ \\
$5.0$   & $0.2\pi$ & \hspace*{3mm}  $1$ & $9502580(1)$ \\
$20.0$  & $0.2\pi$ & \hspace*{3mm}  $9$ & $4111164(1)$ \\
$100.0$ & $0.2\pi$ & \hspace*{3mm} $49$ & $3993176(2)$ \\ 
\hline
$0.5$   & $0.5\pi$ & \hspace*{3mm} $-0$ & $0798648(2)$ \\
$2.0$   & $0.5\pi$ & \hspace*{3mm}  $0$ & $522629(3)$  \\
$5.0$   & $0.5\pi$ & \hspace*{3mm}  $1$ & $9447786(1)$ \\
$20.0$  & $0.5\pi$ & \hspace*{3mm}  $9$ & $387026(2)$  \\
$100.0$ & $0.5\pi$ & \hspace*{3mm} $49$ & $368324(1)$  \\ 
\hline
$0.5$   & $0.8\pi$ & \hspace*{3mm} $-0$ & $0958(3)$    \\
$2.0$   & $0.8\pi$ & \hspace*{3mm}  $0$ & $5105(3)$    \\
$5.0$   & $0.8\pi$ & \hspace*{3mm}  $1$ & $95026(1)$   \\
$20.0$  & $0.8\pi$ & \hspace*{3mm}  $9$ & $4111165(3)$ \\
$100.0$ & $0.8\pi$ & \hspace*{3mm} $49$ & $399318(1)$  \\
\hline \hline
\end{tabular}
\end{table}

Another support for the conjecture~(\ref{:oc}) results from the
counting of the degeneracies of the excited levels (`conformal
towers').  These could in all cases be seen up to the forth or fifth
excitation of the lowest level.  First, there are no indications of
the existence of a second primary field in the operator content, that
is up to the (at least) forth excited level of the relative
ground-state nothing but the expected levels --- the conformal tower
of the ground-state --- has been detected.  On the other hand, the
degeneracies of the conformal towers for the three possible primaries
at \mbox{$c \! = \! \frac{1}{2}$} differ (at the latest) in the third
excitation, therefore the observed degeneracies that complied with
Eq.~(\ref{:oc}) in all examined cases, strongly support the
conjecture.
 
\setcounter{equation}{0}
\section{Conclusions}
 
It was the aim of this article to examine the finite-size scaling
properties of the quantum Ising chain with one `generalized' defect
(\ref{I:gd}), which has been introduced by an arbitrary transformation
in the algebra of observables in the coupling term at one particular
site of the chain.  The finite-size scaling spectrum for the Ising
chain equipped with the most universal defect that maintains its
global ${\bf Z}_{2}$ symmetry was obtained analytically and proven to
be the same as in the case of one suitably chosen `ordinary' defect
and thus could be expressed in terms of irreps of a shifted $u(1)$
Kac-Moody algebra. The result is (see
Eqs.~(\ref{eq:kap})--(\ref{eq:phi})) that for those generalized
defects that do not break the global ${\bf Z}_{2}$ symmetry of the
quantum Ising chain the partition function in the continuum limit is
in fact identical to that for one ordinary defect. The generalization
to an arbitrary number of defects of this form at commensurate
distances obviously is completely analogous to the case of `ordinary'
defects \cite{hps,bcs}.  Recently, a similar observation \cite{wi} has
been made concerning the Ising quantum chain with three spin couplings
\cite{wi} and the Ashkin-Teller model equipped with special defects
\cite{mydip} that break the symmetries in a similar way in both
systems.  This suggests that there is some sort of `universality' for
quantum chains with defects. This means that the partition function of
the system will not depend on the detailed form of the defect, but
only on the symmetries of the chain and the symmetry breaking caused
by the defect.
 
In the remainder of this article another two-parameter family of
generalized defects given by \mbox{$\sigma^{x}_{N+1} = \beta \cdot
\left( \cos(\psi) \sigma^{x}_{1} - \sin(\psi) \sigma^{z}_{1} \right)$}
with \mbox{$\beta \! \in \! {\bf R}-\{0\}$} and \mbox{$\psi \! \in \!
(0,\pi) $}, that destroy the finite-size integrability property of the
quantum Ising chain has been studied by means of numerical finite-size
calculations for chains with up to sixteen sites.  They suggest a
qualitatively different behaviour of the finite-size scaling spectrum,
which appears to depend non-continuously on the parameter $\psi$ of
the defect. In this case a conjecture for the operator content
involving only one primary field of a \mbox{$c\! = \! \frac{1}{2}$} VA
for each choice of the parameters was given and compared to the
results obtained from the numerical diagonalization of the chain. For
some examples of the parameters, this comparison is presented in the
four tables, where in addition numerical estimates for the
non-universal surface energy (cf. Eq.~(\ref{:c})) are given.  For
\mbox{$\psi\! =\! \pi/2$}, the resulting finite-size scaling spectrum
is the same as for mixed BCs, whereas for \mbox{$\psi\! \neq \!
\pi/2$} one obtains the finite-size scaling spectrum of one of the two
sectors of fixed BCs depending on the sign of $\beta \cos(\psi)$ only.
Therefore, the highest weights $\Delta$ of the \mbox{$c\! =\!
\frac{1}{2}$} Virasoro irreps are observed to be \mbox{$\Delta\! =\!
0$} for \mbox{$\beta\cos(\psi)$} (strictly) positive, \mbox{$\Delta\!
=\! 1/16$} in the case \mbox{$\cos(\psi\!) =\! 0\;$}, and finally
\mbox{$\Delta\! =\! 1/2$} for negative values of
\mbox{$\beta\cos(\psi)\;$}.
 
In this article, not all possible generalized defects were
investigated. In principle, one has four real parameters
$e_{1},e_{x},e{y},$ and $e_{z}$ corresponding to a defect of the form
(\ref{I:gd}) for the quantum Ising Hamiltonian~(\ref{e:H}), so the two
types of defects studied in Sec.~3 and Sec.~4 correspond to
\mbox{$e_{1}\! =\! e_{z}\! = \! 0$} and arbitrary $e_{x}$ and $e_{y}$
resp.  \mbox{$e_{1}\! =\! e_{y}\! = \! 0$} and arbitrary values of
$e_{x}$ and $e_{z}$ .  Thus the question, if the finite-size scaling
spectra for the chain with a localized defect different from those
considered herein will show the same behaviour as observed or not,
remains unanswered.  One might guess that this is true, since a
localized defect should not affect the conformal properties except
that the two \mbox{$c\! = \! \frac{1}{2}$} VAs describing the
translational invariant case become coupled and thereby only their
direct sum appears in the spectrum generating algebra. From this one
would expect that for every choice of a localized defect the spectrum
generating algebra is at least a (shifted) $u(1)$ Kac-Moody
algebra. Furthermore the situations of several isolated, but possibly
different, generalized defects or even extended generalized defects
have not been looked upon. Clearing up these points is left to future
investigations.
 
\setcounter{equation}{0}
\section{Acknowledgements}
 
It is a pleasure for the author to thank V.\ Rittenberg for
substantial support as well as P.\ Chaselon for many fruitful
discussions.

\appendix
\renewcommand{\theequation}{\Alph{section}.\arabic{equation}}
\renewcommand{\thesection}{\mbox{Appendix }\Alph{section}}
\setcounter{section}{0}
\setcounter{equation}{0}
\section{\mbox{}}
 
In this place a simple, but maybe non-obvious corollary of linear
algebra shall be proven.  It has been used in Sec.~3 to show that one
can always find a Bogoliubov transformation~(\ref{eq:bt})
diagonalizing the Hamiltonian~(\ref{eq:HT}) to the form~(\ref{eq:HTD})
and which beyond that is compatible with the
conditions~(\ref{eq:U})~and~(\ref{eq:C}).  This was necessary to
guarantee that the new operators defined by Eq.~(\ref{eq:bt}) respect
the fermionic algebra~(\ref{eq:acr}). For this purpose one needs only
Eq.~(\ref{eq:CAC}) as supposition. This enables us to state the
proposition in a more general way.
 
Let $M$ be a hermitian $2N\! \times \! 2N$ matrix
with the property
\[
C M C = M^{\ast} \; \; ,
\]
where $C$ denotes the matrix (\ref{eq:CC})
\[ 
\begin{array}{ccc}
C = \left( \begin{array}{cc} {\bf 0}_{N} & {\bf 1}_{N} \\ 
{\bf 1}_{N} & {\bf 0}_{N} \end{array} \right) & , & C^2 = 
{\bf 1}_{2N} \; \; .
\end{array} 
\]
Then there exists a set ${\cal B}$ of $2N$ (row) vectors ${\cal B} =
\{ \vec{\Theta}_{k} , k\! =\! 1,\ldots , 2N \}$ fulfilling the three
relations
\begin{eqnarray}
\vec{\Theta}_{j} \cdot \vec{\Theta}_k^{\dagger} & = & 
\delta_{j,k} \vspace{4mm} \\
\vec{\Theta}_{k} \cdot M & = & \Lambda_{k} \cdot 
\vec{\Theta}_{k} \; \; , \; \; \Lambda_{k} \in {\bf R} \vspace{4mm} \\
\vec{\Theta}_k^{\ast} \cdot \, C & = & \vec{\Theta}_{k}
\end {eqnarray}
for all $ j,k \in \{ 1,2,\ldots ,2N \}$.
 
Since $M$ is hermitian, there exists an orthonormal set ${\cal
B}^{(0)} = \{ \vec{\Theta}_{k}^{(0)} ,\; k\! =\! 1, \ldots ,2N \}$ of
eigenvectors of $M$. This means that ${\cal B}^{(0)}$ automatically
satisfies the first two relations (A.1) and (A.2) of the proposition.
The proof is based on an iterative prescription how to built from this
set ${\cal B}^{(0)}$ another set ${\cal B}$ of orthonormal
eigenvectors of $M$ that satisfies the additional third condition in
the proposition.
 
Before commencing with that, consider an arbitrary normalized
eigenvector $\vec{\Theta}$ of $M$ belonging to an eigenvalue
$\Lambda$, that is
\[ 
\begin{array}{ccccc}
\vec{\Theta} \cdot M = \Lambda \cdot \vec{\Theta} & , &
\vec{\Theta} \cdot \vec{\Theta}^{\dagger} = 1 & , &
\Lambda \in {\bf R} \; \; .
\end{array} 
\]
It follows
\[ 
\begin{array}{ccl}
( \vec{\Theta}^{\ast} C) \cdot M & = & 
\vec{\Theta}^{\ast} \cdot ( C M C ) \cdot C \vspace{2mm} \\
                                 & = & 
\vec{\Theta}^{\ast} \cdot M^{\ast} \cdot C \vspace{2mm} \\
                                 & = & 
\Lambda \cdot (\vec{\Theta}^{\ast} C) \; \; , \vspace{4mm} \\
(\vec{\Theta}^{\ast} C ) \cdot (\vec{\Theta}^{\ast} C)^{\dagger} 
& = & 1 \; \; ,
\end{array} 
\]
that is, ($\vec{\Theta}^{\ast} C$) is normalized eigenvector of $M$ to
the same eigenvalue $\Lambda$.
 
Now, start with $\vec{\Theta}_{1}^{(0)}$. Then there are two
possibilities: either the two vectors
\mbox{($\vec{\Theta}_{1}^{(0)^{\scs \ast}} \cdot C$)} and
$\vec{\Theta}_{1}^{(0)}$ that are both eigenvectors of $M$ to the
eigenvalue $\Lambda_{1}$ linear depend on each other or not.  In the
first case you have
\[
\vec{\Theta}_{1}^{(0)^{\scs \ast}} \cdot C = 
e^{2i\gamma} \cdot \vec{\Theta}_{1}^{(0)}
\]
with $\gamma \in [ 0 ,\pi )$. The vector $\vec{\Theta}_{1}$ defined by
\[
\vec{\Theta}_{1} = e^{i\gamma} \cdot \vec{\Theta}_{1}^{(0)}
\]
has the properties
\[ 
\begin{array}{ccc}
\vec{\Theta}_{1}^{\ast} \cdot C = \vec{\Theta}_{1} & , &
\vec{\Theta}_{1} \cdot \vec{\Theta}_{1}^{\dagger} = 1 \; \; .
\end{array} 
\]
Take now as the new set ${\cal B}^{(1)} = \{ \vec{\Theta}_{1},
\vec{\Theta}_{k}^{(0)} ,\; k\! =\! 2, \ldots, 2N \}$, which of course
still fulfills (A.1) and (A.2).  In the second case, define two new
vectors $\vec{\Theta}_{1}$ and $\vec{\Theta}_{2}$ by
\[ 
\begin{array}{ccc}
\vec{\Theta}_{1} & = & \frac{1}{\sqrt{2(1+d)}}
\left( e^{i\delta} \vec{\Theta}_{1}^{(0)}
+ e^{-i\delta} \vec{\Theta}_{1}^{(0)^{\scs \ast}} 
\cdot C \right) \vspace{4mm} \\
\vec{\Theta}_{2} & = & \frac{i}{\sqrt{2(1-d)}}
\left( e^{i\delta} \vec{\Theta}_{1}^{(0)}
- e^{-i\delta} \vec{\Theta}_{1}^{(0)^{\scs \ast}} 
\cdot C \right) \; \; ,
\end{array} 
\]
where
\[
(\vec{\Theta}_{1}^{(0)^{\scs \ast}} \cdot C) \cdot \vec{\Theta}_{1}^{(0)^{\scs \dagger}} = d e^{2i\delta}
\]
with $d \in [0,1)$ and $\delta \in [0,\pi )$.  Now choose $2N\! -\! 2$
new vectors $\vec{\Theta}_{k}^{(1)}$ to be orthonormal eigenvectors of
M perpendicular to the subspace spanned by $\vec{\Theta}_{1}$ and
$\vec{\Theta}_{2}$ such that the two sets ${\cal B}^{(0)}$ and ${\cal
B}^{(1)}$, \mbox{${\cal B}^{(1)} =
\{\vec{\Theta}_{1},\vec{\Theta}_{2},\vec{\Theta}_{k}^{(1)},\; k\! =\!
3,\ldots,2N \} $}, span the same space (one can obviously restrict
this orthogonalization procedure to the set of vectors
$\vec{\Theta}_{k}^{(0)}$ that span the eigenspace of $M$ belonging to
the eigenvalue $\Lambda_{1}$). This ensures that the new set ${\cal
B}^{(1)}$ fulfills the relations (A.1) and (A.2) and, in addition,
\[ 
\begin{array}{ccc}
\vec{\Theta}_{j}^{\ast} \cdot C = 
\vec{\Theta}_{j} & , & j \in \{ 1,2 \} \; \; .
\end{array} 
\]
 
By iteration of this process (that means performing successive changes
of basis where in each step one or two additional vectors comply with
Eq.~(A.3)), one finally --- after at most $2N$ steps --- ends up with
a set $\cal B$ that fulfills the proposition.

\appendix
\renewcommand{\theequation}{\Alph{section}.\arabic{equation}}
\renewcommand{\thesection}{\mbox{Appendix }\Alph{section}}
\setcounter{equation}{0}
\setcounter{section}{1}
\section{\mbox{}}

This Appendix contains another proof that has been omitted in Sec.~3
in order to keep the representation easier to survey.  It shall be
shown that all eigenvalues of the matrix $\widehat{A}(\alpha, \phi )$
(\ref{eq:AT}) are at least doubly degenerate (in fact the degree of
degeneration is always even). Even more, if $\widehat{U}$
(\ref{eq:UT}) is any unitary transformation compatible with the
condition~(\ref{eq:UC}) that diagonalizes $\widehat{A}(\alpha ,\phi )$
for some fixed values of the parameters, then the diagonal matrix
\mbox{$\widehat{A}_{D} = \widehat{U}^{\dagger} \widehat{A}(\alpha
,\phi ) \widehat{U}$} has the form
\begin{equation}
\widehat{A}_{D} = \left( 
\begin{array}{cc} \Lambda^{2} & {\bf 0}_{N} \\ 
{\bf 0}_{N} & \Lambda^{2} \end{array} \right) \; \; ,
\label{eq:B1}
\end{equation}
where $\Lambda$ denotes the diagonal $N\! \times \! N$ matrix with
elements $\Lambda_{k} , \; k\! =\! 1, \ldots ,N$.
 
The proof of this fact will be organized as follows. First, we will
take into account that the Bogoliubov transformation~(\ref{eq:bt}) is
of course not uniquely determined by the claimed properties. From this
consideration, it will be possible to find two different
transformations that yield the same diagonal matrix $\widehat{A}_{D}$
and the conjectured form of $\widehat{A}_{D}$ follows directly. As
will be seen in what follows, the real cause of the degeneration lies
in the fact that by exchanging the role of creation and annihilation
operators one changes the sign of the $\Lambda_{k}$.
 
Assume that one succeeded to find a certain unitary transformation
$\widehat{U}$ that diagonalizes \mbox{$\widehat{A}(\alpha , \phi )$}
for definite values of $\alpha$ and $\phi$ and in addition complies
with the condition~(\ref{eq:UC}). It was proved in Appendix~A that
this is always possible. Then you have
\begin{equation}
\widehat{A}_{D}(\alpha , \phi ) = 
\widehat{U}^{\dagger} \widehat{A}(\alpha ,\phi ) \widehat{U}
= \left( \begin{array}{cc} \Lambda^{2} & {\bf 0}_{N} \\ 
{\bf 0}_{N} & \Lambda^{\prime^{\scs 2}} \end{array} \right) \; \; ,
\label{eq:B5}
\end{equation}
where now $\Lambda$ and $\Lambda^{\prime}$ are diagonal $N\! \times \!
N$ matrices with elements $\Lambda_{k}$ and $\Lambda^{\prime}_{k}$,
respectively \mbox{($k\! =\! 1,\ldots , N$)}.  Now, suppose you
perform a Bogoliubov transformation (\ref{eq:bt}) involving only
creation and annihilation operators $a_{j}^{\dagger}$ and $a_{j}$ with
one fixed index $j$, that is
\begin{equation} 
\begin{array}{ccl}
\tilde{a}_{j}           & = & 
g a_{j} + h^{\ast} a_{j}^{\dagger} \vspace{4mm} \\
\tilde{a}_{j}^{\dagger} & = & 
h a_{j} + g^{\ast} a_{j}^{\dagger} \; \; ,
\label{eq:bts}
\end{array} 
\end{equation}
where the $2\! \times \! 2$ matrix $u$ defined by
\[
u = \left( \begin{array}{cc} g & h^{\ast} \\ h & g^{\ast} \end{array}
\right)
\]
is unitary (note that this implies that either $g$ or $h$ is equal to
zero and that the determinant $\det (u) \! \in \! \{ 1,-1 \} $).  This
transformation will not alter the anticommutation relations, thus the
operators $\tilde{a}_{j}^{\dagger}$ and $\tilde{a}_{j}$ are fermionic
operators as well.  On the other hand, the Hamiltonian $\tilde{H}$
(\ref{eq:HTD}) given by
\begin{equation}
\tilde{H} = \sum_{k=1}^{N} \Lambda_{k} a_{k}^{\dagger} a_{k} + E_{0}
\label{eq:B2}
\end{equation}
is transformed into
\begin{equation}
\tilde{H} = \sum_{k=1}^{N} (1\! -\! \delta_{j,k}) 
   \Lambda_{k} a_{k}^{\dagger} a_{k} \;
   + \det (u) \Lambda_{j} \tilde{a}_{j}^{\dagger} \tilde{a}_{j}
   + \Lambda_{j} h h^{\ast} + E_{0} \; \; .
\label{eq:B3}
\end{equation}
This means that if $\det (u) \! =\! 1$ the Hamiltonian is not changed
(since this implies $h\! =\! 0$), whereas in the case $\det (u) \! =\!
-1$ the sign of $\Lambda_{j}$ is altered and the ground-state energy
is shifted by the amount \mbox{$\Lambda_{j} h h^{\ast} \! =\!
\Lambda_{j}$}.  Of course one can perform a transformation as such
independently for every value of \mbox{$k,\; k\! =\! 1,\ldots ,N$}.
 
It follows that the equations deduced from the necessary
conditions~(\ref{eq:aH})
\begin{equation}
[ a_{k} , \tilde{H} ] = \Lambda_{k} a_{k}
\label{eq:B4}
\end{equation}
are --- up to the signs of the $\Lambda_{k}$ --- identical,
independent of any performed transformation of the above-mentioned
type. The shift in the constant term $E_{0}$ (see
Eqs.~(\ref{eq:B2})~and~(\ref{eq:B3})) drops out because of the use of
the commutator in Eq.~(\ref{eq:B4}).  However, in the final eigenvalue
equations (\ref{eq:L2}) only the squares of the $\Lambda_{k}$ enter
the calculation, therefore these are not altered at all. This leads to
the consequence that the diagonal matrix $\widehat{A}_{D}$ is
invariant under any of these transformations (\ref{eq:bts}), where of
course one has to take into account the change of basis performed in
the derivation of the eigenvalue equation.  In other words, the whole
problem has a $U(1)^{\otimes N}$-invariance corresponding to an
arbitrary phase factor for each pair of creation and annihilation
operators.  But the eigenvalue equations~(\ref{eq:L2}) beyond this are
not altered by an exchange of creation and annihilation operators,
therefore they enjoy an $\left( U(1) {\otimes}_{s} {\bf Z}_{2} \right)
^{\otimes N}$-invariance, where ${\otimes}_{s}$ denotes a semi-direct
product with the normal divisor (invariant subgroup) on the left side.
 
Now, take as a special transformation of this form the matrix
$\widehat{C}$ defined by
\[
\widehat{C} = \left( \begin{array}{cc} {\bf 0}_N & -i{\bf 1}_{N} \\ 
i{\bf 1}_{N} & {\bf 0}_{N} \end{array} \right) \; \; .
\]
This acts on the $a_{k}^{\dagger}$ and $a_{k}$ in the same way for all
$k$, namely through
\[ 
\begin{array}{ccccc}
a_{k}           & \longmapsto & \tilde{a}_{k} & = &
 i a_{k}^{\dagger}   \vspace{4mm} \\
a_{k}^{\dagger} & \longmapsto & \tilde{a}_{k}^{\dagger} & = &
 -i a_{k} \; \; .
\end{array} 
\]
It follows that the transformation $V$ that is related to $\widehat{U}
= T U$ \mbox{(see Eqs.~(\ref{eq:T})~and~(\ref{eq:UT}))} by
\[
V = T \widehat{C} U = C T U = C \widehat{U} = C \widehat{U}^{\ast} C
\]
with
\[ 
\begin{array}{ccc}
C = \left( \begin{array}{cc} {\bf 0}_{N} & {\bf 1}_{N} \\ 
{\bf 1}_{N} & {\bf 0}_{N} \end{array} \right) & , &
T = \frac{1}{\sqrt{2}} \left( \begin{array}{cc} {\bf 1}_{N} & 
{\bf 1}_{N} \\ i {\bf 1}_{N} & -i {\bf 1}_{N} \end{array} \right) \; \; ,
\end{array} 
\]
diagonalizes $\widehat{A}$ to the same diagonal matrix
$\widehat{A}_{D}$ as $\widehat{U}$ does.  That means that
$\widehat{A}_{D}$ has to commute with $C$ and, using
Eq.~(\ref{eq:B5}), one therefore obtains the result
\[
\Lambda^{2} = \Lambda^{\prime^{\scs 2}}
\]
that proves the assertion~(\ref{eq:B1}).

\end{document}